\documentclass[twocolumn]{aa}
\usepackage{graphicx}
\usepackage{txfonts}
\begin{document}
\title{High mass of the type IIP supernova 2004et inferred from hydrodynamic
       modeling}

\author{V. P. Utrobin\inst{1,2} \and N. N. Chugai\inst{3}}

\offprints{V. Utrobin, \email{utrobin@itep.ru}}

\institute{
   Max-Planck-Institut f\"ur Astrophysik,
   Karl-Schwarzschild-Str. 1, D-85741 Garching, Germany
\and
   Institute of Theoretical and Experimental Physics,
   B.~Cheremushkinskaya St. 25, 117218 Moscow, Russia
\and
   Institute of Astronomy of Russian Academy of Sciences,
   Pyatnitskaya St. 48, 109017 Moscow, Russia}

\date{Received 3 April 2009 / accepted 7 July 2009}

\abstract{
Previous studies of type IIP supernovae have inferred that progenitor masses recovered
   from hydrodynamic models are higher than $15~M_{\sun}$.
}{
To verify the progenitor mass of this supernova category, we attempt
   a parameter determination of the well-observed luminous type IIP supernova
   2004et.
}{
We model the bolometric light curve and the photospheric velocities of SN~2004et
   by means of hydrodynamic simulations in an extended parameter space.
}{
From hydrodynamic simulations and observational data, we infer a presupernova
   radius of $1500\pm140 R_{\sun}$, an ejecta mass of $24.5\pm1 M_{\sun}$,
   an explosion energy of $(2.3\pm0.3)\times10^{51}$ erg, and a radioactive
   $^{56}$Ni mass of $0.068\pm0.009 M_{\sun}$.
The estimated progenitor mass on the main sequence is in the range of
   $25-29 M_{\sun}$.
In addition, we find clear signatures of the explosion asymmetry in the nebular
   spectra of SN~2004et.
}{
The measured progenitor mass of SN~2004et is significantly higher than the
   progenitor mass suggested by the pre-explosion images.
We speculate that the mass inferred from hydrodynamic modeling is overestimated
   and crucial missing factors are multi-dimensional effects.
}
%
\titlerunning{Progenitor of SN~2004et}
\authorrunning{V. P. Utrobin \& N. N. Chugai}
\maketitle

\section{Introduction}
\label{sec:intro}
The major parameters of core-collapse supernova (SN) are thought to be
   linked to the initial stellar mass on the main sequence,
   the progenitor mass.
However, the genealogy of different varieties of SNe is as yet poorly known.
Fortunately, hydrodynamic modeling of the light curves and the expansion
   velocities allows us to estimate SN parameters such as a pre-SN radius,
   an ejecta mass, an explosion energy, and a radioactive $^{56}$Ni amount.
In the case of SNe~IIP, the mass lost prior to the pre-SN stage is
   relatively small for stars with an initial mass less than $25~M_{\sun}$,
   so the ejecta mass provides us with a reliable estimate of the progenitor
   mass or at least its lower limit.
Compared to other core-collapse SNe~Ib/c and SNe~IIn, we are able to recover the ejecta
   mass of SNe~IIP from hydrodynamic modeling with greater confidence because of
   both an accurate estimation of the photospheric velocity related to the high
   opacity of the hydrogen-rich matter, and a low contribution of the
   circumstellar interaction to the SN luminosity.

Hydrodynamic modeling should only be applied to well-observed
   SNe~IIP.
An adequate simulation of the bolometric light curve and the evolution in the
   photospheric velocities requires both high-quality photometric and
   spectroscopic data (Utrobin \cite{Utr_07}).
For the above reason, hydrodynamic simulations of SNe~IIP have been
   performed for only a handful of events: SN~1987A, SN~1999em, SN~2003Z, and
   SN~2005cs.
For these particular cases, the inferred
   progenitor masses have been found unexpectedly to be in the range of $15-22~M_{\sun}$
   (Utrobin \& Chugai \cite{UC_08}), well above the median value of
   $\approx13~M_{\sun}$ for the Salpeter initial mass distribution in
   the mass range of $9-25~M_{\sun}$ responsible for SNe~IIP (Heger et al.
   \cite{HFWLH_03}),
i.e., the progenitor masses are on average more
   massive than expected.

A more challenging problem is one related to the analysis of the
   sub-luminous SN~2005cs.
The progenitor mass of $\sim~18~M_{\sun}$ inferred
   from hydrodynamic modeling (Utrobin \& Chugai \cite{UC_08})
   was found to significantly exceed the progenitor mass of $6-13~M_{\sun}$ recovered
   from archival \emph{HST\/} images (Maund et al. \cite{MSD_05};
   Li et al. \cite{LVF_06}; Eldridge et al. \cite{EMS_07}).
No reasonable explanation of this disparity has been proposed.

To pinpoint the cause of this disagreement between the two methods of the mass
   determination, one needs to confirm this discrepancy in mass measurements
   for a larger sample of SNe~IIP with a broad range of observational
   characteristics.
In this respect, the well-observed luminous SN~2004et in the nearby galaxy
   NGC 6946 is a particularly favorable case.
This object discovered soon after its explosion has the highest
   intrinsic luminosity among well-studied events (Sahu et al. \cite{SASM_06})
   and, perhaps, the highest ejecta mass and explosion energy.
In this case, the progenitor was directly identified in the archival
   images (Li et al. \cite{LVFC_05}).

Here we perform hydrodynamic modeling of SN~2004et to recover the
   basic parameters: pre-SN radius, ejecta mass, explosion energy, and
   radioactive $^{56}$Ni mass.
A brief description of the hydrodynamic model is given in
   Sect.~\ref{sec:modprog-oview}, and the basic parameters of the optimal
   model are obtained in Sect.~\ref{sec:modprog-param}.
In Sect.~\ref{sec:modprog-evpresn}, we investigate whether the
   non-evolutionary model should be used instead of evolutionary pre-SN for
   the one-dimensional hydrodynamic modeling of SN~2004et and in general SNe~IIP.
The progenitor mass of SN~2004et is evaluated and compared to estimations
   for other SNe~IIP (Sect.~\ref{sec:modprog-prgmass}).
The measured progenitor mass noticeably exceeds the mass estimated from the
   pre-explosion images, and this disagreement is discussed in
   Sect.~\ref{sec:disc-prob}.
In particular, we propose that the explosion asymmetry could be responsible for
   the disagreement in mass estimates and explore signatures of the explosion asymmetry
   in the SN~2004et nebular spectra (Sect.~\ref{sec:disc-asym}).
Finally, in Sect.~\ref{sec:concl}, we summarize the results obtained.

We adopt a distance to NGC 6946 of 5.5 Mpc and a reddening $E(B-V)=0.41$
   as measured by Li et al. (\cite{LVFC_05}), an explosion date on September
   22.0 UT (JD 2453270.5), and a recession velocity to the host galaxy of 45
   km\,s$^{-1}$ following Sahu et al. (\cite{SASM_06}).

\section{Hydrodynamic model and progenitor mass}
\label{sec:modprog}

\subsection{Model overview}
\label{sec:modprog-oview}
The modeling of the SN explosion is performed using the spherically-symmetric
   hydrodynamic code with one-group radiation transfer (Utrobin \cite{Utr_04},
   \cite{Utr_07}), which has been applied previously to other SNe~IIP.
Utrobin (\cite{Utr_07}) found that both this one-group approach and the multi-group
   approach of Baklanov et al. (\cite{BBP_05}) measured similar ejecta mass
   and explosion energy for SN~1999em.
The basic equations and details of the input physics, including calculations of
   mean opacities, are described in Utrobin (\cite{Utr_04}).
The present version of the code includes additionally Compton cooling and
   heating.
The explosion energy is modeled by placing the supersonic piston close to the outer
   edge of the 1.6 $M_{\sun}$ central core, which is removed from the computational
   mass domain and assumed to collapse to become a neutron star.
The principal limitation of the code is that the explosion asymmetry and the
   Rayleigh-Taylor mixing between the helium core and hydrogen envelope
   (M\"{u}ller et al. \cite{MFA_91}) cannot be correctly treated by the
   one-dimensional model.
We, therefore, study a "non-evolutionary" pre-SN, which takes into account
   the result of the mixing during the explosion and the shock
   propagation in the evolutionary pre-SN.
The distinctive feature of the non-evolutionary model is a smoothed density
   and composition jumps between the helium core and the hydrogen envelope.

\begin{figure}[t]
   \resizebox{\hsize}{!}{\includegraphics[clip, trim=0 0 0 20]{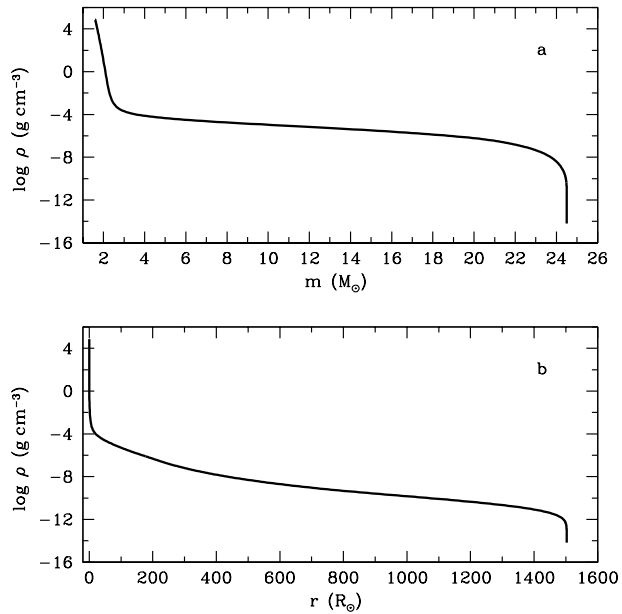}}
   \caption{%
   Density distribution as a function of interior mass \textbf{a}) and
   radius \textbf{b}) for the optimal pre-SN model.
   The central core of 1.6 $M_{\sun}$ is omitted.
   }
   \label{fig:denmr}
\end{figure}
\begin{figure}[b]
   \resizebox{\hsize}{!}{\includegraphics[clip, trim=0 0 0 240]{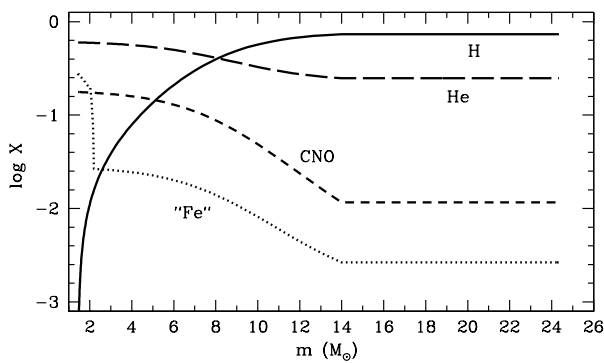}}
   \caption{%
   The mass fraction of hydrogen (\emph{solid line\/}), helium
      (\emph{long dashed line\/}), CNO elements (\emph{short dashed line\/}),
      and Fe-peak elements including radioactive $^{56}$Ni
      (\emph{dotted line\/}) in the ejecta of the optimal model.
   }
   \label{fig:chcom}
\end{figure}

The resultant structure of the non-evolutionary pre-SN  in our optimal model
   is shown in Fig.~\ref{fig:denmr}.
The pre-SN model is defined to be a red supergiant (RSG) with a radius of
   $1500~R_{\sun}$, three times larger than in the case of the normal type IIP
   SN~1999em (Utrobin \cite{Utr_07}).
This large pre-SN radius for SN~2004et is implied by the broad initial peak of
   the bolometric light curve shown by Sahu et al. (\cite{SASM_06}).
The adopted mixing between the helium core and hydrogen envelope in the optimal
   model is shown in Fig.~\ref{fig:chcom}.
The degree of mixing determines the light curve at the end of the plateau
   (Utrobin et al. \cite{UCP_07}).
The unmixed helium-core mass adopted for SN~2004et is $8.1~M_{\sun}$, which
   corresponds to the final helium core of a main-sequence star of
   $\approx25~M_{\sun}$ (Hirschi et al. \cite{HMM_04}).
We note that the model light curve is not sensitive to any variation in the
   helium-core mass of the mixed model (Utrobin et al. \cite{UCP_07}).
%

\subsection{Basic parameters}
\label{sec:modprog-param}
The search for the best-fit model parameters is performed by computations of
   an extensive grid of hydrodynamic models.
The optimal model should reproduce simultaneously the bolometric light curve and
   the photospheric velocity evolution in the best way.
As a result, we derive the ejecta mass $M_{env}=22.9\pm1~M_{\sun}$, the explosion
   energy $E=(2.3\pm0.3)\times10^{51}$ erg, the pre-SN radius
   $R_0=1500\pm140~R_{\sun}$, and the $^{56}$Ni mass
   $M_{\mathrm{Ni}}=0.068\pm0.009~M_{\sun}$.
The uncertainties in the basic parameters are calculated by assuming relative errors
   in the input observational data: 11\% in the distance, 7\% in the dust
   absorption, 5\% in the photospheric velocity, and 2\% in the plateau duration.
In general, the errors of derived parameters should be somewhat larger because of
   model systematic errors.
However, the latter cannot be confidently estimated unless a more advanced and
   correct model is developed.
The model uncertainties will be discussed below in Sect.~\ref{sec:disc-asym}.

\begin{figure}[t]
   \resizebox{\hsize}{!}{\includegraphics[clip, trim=0 0 0 120]{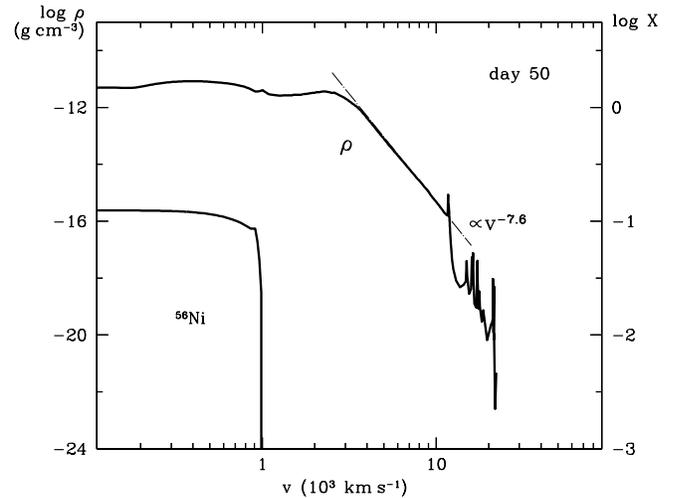}}
   \caption{%
   The density and the $^{56}$Ni mass fraction as a function of velocity
      for the optimal model at $t=50$ days.
   \emph{Dash-dotted line} is the density distribution fit
      $\rho \propto v^{-7.6}$.
   }
   \label{fig:denicl}
\end{figure}
The density distribution in the freely expanding SN envelope is shown in
   Fig.~\ref{fig:denicl}.
Multiple shells in the outermost layers with velocities $v>14\,000$ km\,s$^{-1}$
   (Fig.~\ref{fig:denicl}) form at the shock breakout stage by the radiative
   acceleration in the optically thin regime.
The origin of these shells is related to the specific behavior of the line
   opacity in the outer rarefied layers of temperature $\sim 10^5$~K.
The innermost shell of mass $\sim5\times10^{-3}~M_{\sun}$ at the
   velocity of 11\,700 km\,s$^{-1}$ is the thin shell formed by the effect
   of the shock breakout in the optically thick regime
   (Grassberg et al. \cite{GIN_71}; Chevalier \cite{Che_81}).

\begin{figure}[b]
   \resizebox{\hsize}{!}{\includegraphics[clip, trim=0 0 0 240]{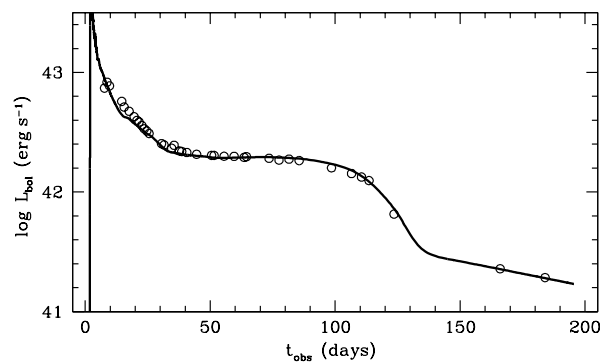}}
   \caption{%
   The calculated bolometric light curve of the optimal model
      (\emph{solid line\/}) overplotted on the bolometric data of SN 2004et
      evaluated from the $UBVRI$ observations of Sahu et al.
      (\cite{SASM_06}) (\emph{open circles\/}).
   }
   \label{fig:lmbol}
\end{figure}
The optimal model describes the bolometric light curve quite well,
   including its initial ($t<30$ days) peak (Fig.~\ref{fig:lmbol}).
This peak is substantially broader and more luminous than the initial peak
   of SN~1999em (cf. Sahu et al. \cite{SASM_06}).
It is the initial luminosity peak of SN~2004et that requires the larger radius
   of the pre-SN model compared to the pre-SN radius of $500~R_{\sun}$ in the
   case of SN~1999em (Utrobin \cite{Utr_07}).

\begin{figure}[t]
   \resizebox{\hsize}{!}{\includegraphics[clip, trim=0 0 0 120]{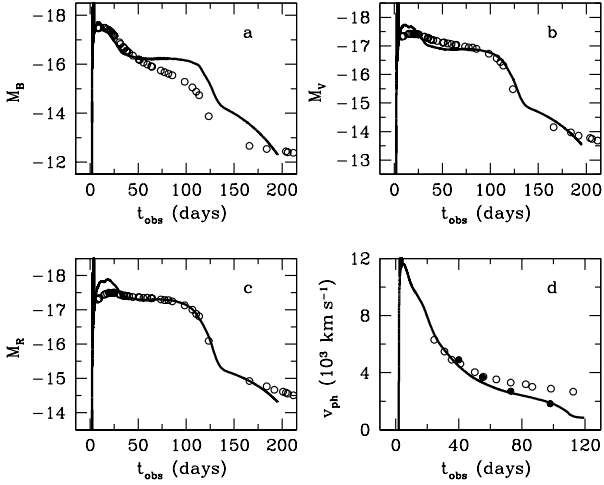}}
   \caption{%
   Optimal hydrodynamic model.
   Panels \textbf{a}), \textbf{b}), and \textbf{c}): the calculated $B$, $V$, and $R$
      light curves (\emph{solid lines\/}) compared with the corresponding observations
      of SN 2004et obtained by Sahu et al. (\cite{SASM_06}) (\emph{open circles\/}).
   Panel \textbf{d}): the calculated photospheric velocity (\emph{solid
      line\/}) is compared with photospheric velocities estimated from the
      absorption minimum of the Fe\,II 5169 \AA\ line (\emph{open circles\/})
      by Sahu et al. (\cite{SASM_06}) and recovered from the Na\,I doublet profile
      (\emph{filled circles\/}).
   }
   \label{fig:optmod}
\end{figure}
With one-group radiation transfer, the hydrodynamic model is not assumed to
   reproduce the monochromatic light curves in detail.
However, it is instructive to compare the model and the observations in $B$, $V$,
   and $R$ bands (Figs.~\ref{fig:optmod}a,b,c).
The $B$ light curve is reproduced at the initial hot phase, but not at the
   late phase.
The overall fit of the $V$ light curve is much tighter.
In $R$ band, the calculated light curve reproduces the plateau data, but does not
   describe the very initial stages of the light curve.
The differences between the hydrodynamic model and the observations are related
   to deviations of the SN spectrum from a blackbody.
These deviations are significant in the blue and ultraviolet at late
   photospheric epochs, which explains why the disagreement is strongest
   in $B$ band.

The computed photospheric velocity is shown together with two sets of
   observational data (Fig.~\ref{fig:optmod}d): the first is recovered from an
   absorption minimum of the Fe\,II~5169 \AA\ line (Sahu et al. \cite{SASM_06})
   and the second, from our modeling of the Na\,I doublet profile.
The latter photospheric velocities can be measured more confidently than those
   recovered from absorption minima, especially at late photospheric epochs
   when absorption lines become strong.
The model photospheric velocity is consistent with both the Na\,I data and the early
   data of the Fe\,II~5169 \AA\ absorption minimum.
We also modeled profiles of the Fe\,II 4924, 5018, 5169 \AA\ lines at late
   photospheric stages, and found that the photospheric velocities obtained from
   these lines were rather similar to velocities obtained by modeling
   the Na\,I doublet profile.
Unfortunately, the spectral data of SN 2004et for the early stages are missing.
The blue edge of the H$\alpha$ absorption in the earliest spectrum on day~24
   implies a maximal expansion velocity of $\sim12\,000-13\,000$ km\,s$^{-1}$
   in the ejecta.
This velocity is consistent with the model maximal velocity of 12\,000 km\,s$^{-1}$.

\subsection{Explosion of evolutionary presupernova}
\label{sec:modprog-evpresn}
%
\begin{figure}[b]
   \resizebox{\hsize}{!}{\includegraphics[clip, trim=0 220 0 120]{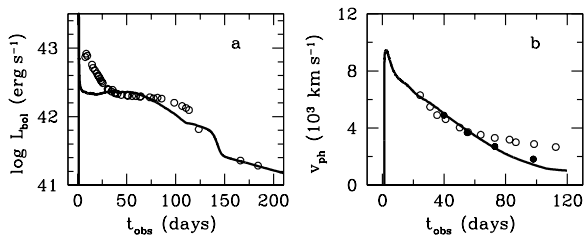}}
   \caption{%
   Hydrodynamic model of SN 2004et for the evolutionary pre-SN.
   Panel \textbf{a}): the model bolometric light curve (\emph{solid line\/})
      overplotted on the observational data (see Fig.~\ref{fig:lmbol}
      legend for details).
   Panel \textbf{b}): the model photospheric velocity (\emph{solid line\/})
      and the observational photospheric velocities (see Fig.~\ref{fig:optmod}d
      legend for details).
   }
   \label{fig:evmod}
\end{figure}
The arguments in favor of the non-evolutionary pre-SN model leave some doubts
   and raise a question: why should we not consider an evolutionary pre-SN?
This issue has already been explored for the sub-luminous type IIP SN~2005cs
   (Utrobin \& Chugai \cite{UC_08}), for which we found that the evolutionary
   pre-SN did not allow us to produce a realistic description of both the light curve
   and the photospheric velocities.
A similar problem was encountered while modeling the explosion of evolutionary pre-SNe with
   other hydrodynamic codes (Chieffi et al. \cite{CDH_03}; Woosley \& Heger
   \cite{WH_07}).

We therefore check whether the same problem holds for SN~2004et, which
   differs in terms of both ejecta mass and explosion energy from SN~2005cs.
We adopt the pre-SN model with an envelope mass of $15.9~M_{\sun}$
   and a density distribution that closely resembles that of the evolutionary model (Utrobin \&
   Chugai \cite{UC_08}).
The hydrogen and helium are assumed to be mixed along the mass coordinate
   in a similar way to the optimal model (Fig.~\ref{fig:chcom}).
An optimal fit to the bolometric light curve and the evolution
   in photospheric velocity is attained for an explosion energy of
   $1.3\times10^{51}$ erg and a pre-SN radius of $600~R_{\sun}$.
Apparent disadvantages of the obtained model are an extremely narrow initial
   peak of luminosity and a two step-like transition from the plateau to the
   radioactive tail (Fig.~\ref{fig:evmod}a).
The latter behavior is similar to that demonstrated by the light curves
   computed for the evolutionary pre-SN by Woosley \& Heger (\cite{WH_07}).
The narrow initial peak is related to the relatively small pre-SN radius.
However, one cannot increase the pre-SN radius to obtain a superior fit because
   the photospheric velocity would then become unacceptably low.
A larger initial radius would produce a higher luminosity,
   which, in turn, should be compensated by a decrease in the explosion
   energy, consequently, leading to lower expansion velocities.
Even in the demonstrated model, the maximal velocity is only 9500 km\,s$^{-1}$
   (Fig.~\ref{fig:evmod}b), significantly lower than the observed maximal
   velocity of $12\,000-13\,000$ km\,s$^{-1}$.
We also computed the same model but without mixing between the helium core and
   the hydrogen-rich envelope.
This model provides an even poorer fit because the "bump" at the end of the plateau
   becomes more boxy, in sharp contrast to the observational light curve.

To summarize, the model including an explosion of the evolutionary pre-SN does not allow us to
   achieve a close fit to the bolometric light curve and the maximal
   expansion velocities of SN~2004et.
This problem is not related to the evolutionary pre-SN itself.
The one-dimensional hydrodynamics cannot reproduce
   the outcome of a real explosion in the evolutionary model, because
   multi-dimensional effects, in particular mixing between the helium core and
   the hydrogen envelope, play a crucial role during the explosion and shock
   propagation phases.
The two-dimensional hydrodynamic model predicts that the Rayleigh-Taylor
   mixing at the helium/hydrogen interface reduces the high
   composition and density gradients (M\"{u}ller et al. \cite{MFA_91}).
Our non-evolutionary pre-SN qualitatively takes this multi-dimensional effect
   into account, which ensures that the non-evolutionary model describes
   the light curve shape more successfully.
A non-evolutionary pre-SN is also preferred by Baklanov et al. (\cite{BBP_05})
   in their modeling of SN~1999em.

\subsection{Progenitor mass}
\label{sec:modprog-prgmass}
The ejecta mass $M_{env}=22.9\pm1~M_{\sun}$ combined with the collapsing core of
   $1.6~M_{\sun}$ yields the pre-SN mass of $24.5\pm1~M_{\sun}$.
A progenitor mass on the main sequence is higher by the amount of matter lost
   via the wind at the main-sequence and RSG phases.
For the main-sequence stage, we rely on the computations of Meynet \& Maeder
   (\cite{MM_03}) for non-rotating stars with the mass-loss rate of Vink
   et al. (\cite{VKL_01}).
They found that a star with an initial mass $M_{ZAMS}=25~M_{\sun}$ lost
   $0.8~M_{\sun}$ during the main sequence (Meynet \& Maeder \cite{MM_03}).
For the RSG stage, Meynet \& Maeder (\cite{MM_03}) used the mass-loss rate of
   de Jager (\cite{JNH_88}) and predicted that $25~M_{\sun}$ and
   $40~M_{\sun}$ main-sequence stars lost $7.5~M_{\sun}$ and $21~M_{\sun}$,
   respectively.
With these estimates of the lost mass, we come to the progenitor mass of SN~2004et
   in the range of $30-40~M_{\sun}$.

A less massive progenitor is expected if we use the wind density recovered for
   the SN~2004et pre-SN from X-ray observations.
These data suggest the mass-loss rate of $(2-2.5)\times10^{-6}~M_{\sun}$\,yr$^{-1}$,
   assuming the wind velocity of 10 km\,s$^{-1}$ (Rho et al. \cite{RJCC_07};
   Misra et al. \cite{MPC_07}).
Using the RSG life-time of $7\times10^5$ yr for the $25~M_{\sun}$ main-sequence
   star (Hirshi et al. \cite{HMM_04}), we find the mass lost at the RSG phase
   to be $\sim1.6~M_{\sun}$ with an uncertainty of $\sim\pm1~M_{\sun}$.
With the mass of $0.8~M_{\sun}$ lost at the main-sequence phase (Hirshi et al.
   \cite{HMM_04}), the total lost mass is then $2.4\pm1~M_{\sun}$.
The pre-SN mass of $24.5~M_{\sun}$ combined with the lost mass results in
   the progenitor mass of $27\pm2~M_{\sun}$, where the error includes
   the uncertainties in the ejecta mass and the mass-loss rate.
The progenitor mass of SN~2004et turns out to be close to the maximal initial
   mass for SNe~IIP according to the present-day paradigm (Heger et al.
   \cite{HFWLH_03}).

\begin{figure}[t]
   \resizebox{\hsize}{!}{\includegraphics[clip, trim=0 0 0 20]{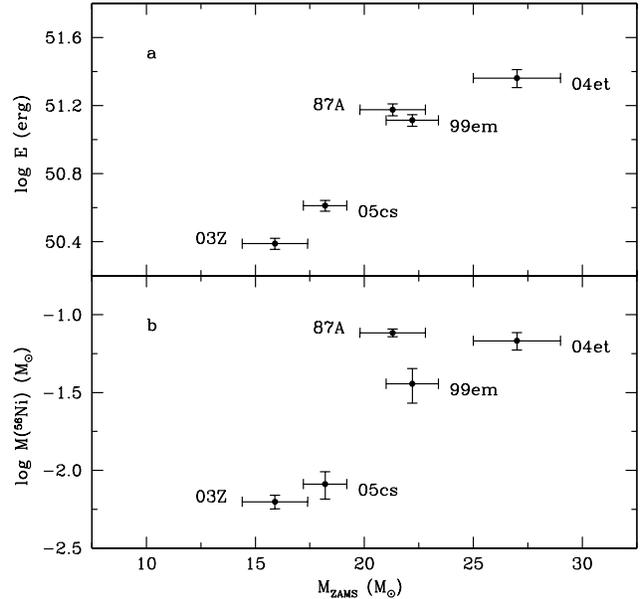}}
   \caption{%
   Explosion energy \textbf{a}) and $^{56}$Ni mass \textbf{b}) versus
      hydrodynamic progenitor mass for five core-collapse SNe.
   }
   \label{fig:nienms}
\end{figure}

\begin{table}[t]
\caption[]{Hydrodynamic models of type IIP supernovae.}
\label{tab:sumtab}
\centering
\begin{tabular}{ l c c c c c c }
\hline\hline
\noalign{\smallskip}
 SN & $R_0$ & $M_{env}$ & $E$ & $M_{\mathrm{Ni}}$ 
       & $v_{\mathrm{Ni}}^{max}$ & $v_{\mathrm{H}}^{min}$ \\
       & $(R_{\sun})$ & $(M_{\sun})$ & ($10^{51}$ erg) & $(10^{-2} M_{\sun})$
       & \multicolumn{2}{c}{(km\,s$^{-1}$)}\\
\noalign{\smallskip}
\hline
\noalign{\smallskip}
 1987A &  35  & 18   & 1.5   & 7.65 &  3000 & 600 \\
1999em & 500  & 19   & 1.3   & 3.60 &  660  & 700 \\
 2003Z & 229  & 14   & 0.245  & 0.63 &  535  & 360 \\
2004et & 1500 & 22.9 & 2.3   & 6.8  &  1000 & 300 \\
2005cs & 600  & 15.9 & 0.41  & 0.82 &  610  & 300 \\
\noalign{\smallskip}
\hline
\end{tabular}
\end{table}
Table~\ref{tab:sumtab} presents the parameters of all the SNe~IIP studied
   hydrodynamically.
The listed parameters are the pre-SN radius, the ejecta mass, the explosion
   energy, the total $^{56}$Ni mass, the maximal velocity of $^{56}$Ni mixing
   zone, and the minimal velocity of the hydrogen-rich envelope.
The ejecta and progenitor masses of SN~2004et are found to be maximal among the
   well-studied SNe~IIP.
With the exception of the initial radius for SN~1987A, the explosion energy and
   the total $^{56}$Ni mass show the most extreme variations (of one order of
   magnitude).
All SNe~IIP are characterized by a deep mixing of hydrogen, indicated by
   the low value of $v_{\mathrm{H}}^{min}$, which is consistent with
   two-dimensional simulations (M\"{u}ller et al. \cite{MFA_91};
   Kifonidis et al. \cite{KPSJM_03}, \cite{KPSJM_06}).
The position of SN~2004et on the plots of explosion energy versus
   progenitor mass (Fig.~\ref{fig:nienms}a) and the total $^{56}$Ni mass
   versus the progenitor mass (Fig.~\ref{fig:nienms}b) strengthens
   the correlations recovered earlier for the SNe~IIP studied hydrodynamically
   (Utrobin \& Chugai \cite{UC_08}).
We note that these correlations also infer the correlation between
   the explosion energy and the total $^{56}$Ni mass, which was found and
   discussed earlier by Nadyozhin (\cite{Nad_03}).
%

\section{Discussion}
\label{sec:disc}

\subsection{Whether the hydrodynamic mass is overestimated?}
\label{sec:disc-prob}
The pre-SN mass recovered from the hydrodynamic modeling is very much close to
   the progenitor mass.
It is, therefore, reasonable to refer to the progenitor mass thus determined as
   the "hydrodynamic" mass.
Surprisingly, the hydrodynamic mass of the SN~2004et progenitor,
   $27\pm2~M_{\sun}$, noticeably exceeds, by a factor of $\sim2-3$,
   the value of $9^{+5}_{-1}~M_{\sun}$ recovered from the analysis of
   pre-explosion images (Smartt et al. \cite{SECM_09}).
A similarly large mass disparity has been found for SN~2005cs: $18~M_{\sun}$
   (Utrobin \& Chugai \cite{UC_08}) versus $6-13~M_{\sun}$
   (Maund et al. \cite{MSD_05}; Li et al. \cite{LVF_06};
   Eldridge et al. \cite{EMS_07}).
These two cases of huge discrepancies in progenitor mass clearly illustrate
   the high level of uncertainty in the mass problem.

The other side of this problem is that all the known hydrodynamic masses of
   SNe~IIP progenitors are in the range $M_{ZAMS}>15~M_{\sun}$
   (Fig.~\ref{fig:nienms}).
This mass distribution conflicts with the paradigm that SNe~IIP
   originate from the mass range of $9-25~M_{\sun}$ (Heger et al.
   \cite{HFWLH_03}).
Assuming the Salpeter initial mass function for SNe~IIP rate in
   the $9-25~M_{\sun}$ mass range and neglecting selection effects,
   we expect that five SNe should be found in the $15-25~M_{\sun}$ mass range with
   a probability of only 0.004.

A quite different conclusion was reached by Smartt et al.
   (\cite{SECM_09}).
They conclude that SNe~IIP progenitors detected or undetected in pre-explosion
   images have masses in the range of $8-17~M_{\sun}$.
This significant difference in the progenitor-mass distributions obtained by
   two methods and the mass discrepancy for SN~2004et and
   SN~2005cs again imply that the determination of the progenitor mass
   remains a difficult problem.
Although the low masses of SNe~IIP progenitors from archival images also pose
   a problem for the fate of massive RSG stars in the range of $17-25~M_{\sun}$
   (Smartt et al. \cite{SECM_09}), we address here only the possibility that
   hydrodynamic masses are strongly overestimated compared to the real
   progenitor masses.

\subsection{Could explosion asymmetry be a crucial missing factor?}
\label{sec:disc-asym}
%
\begin{table}[b]
\caption[]{Components of oxygen line decomposition.}
\label{tab:oxygen}
\centering
\begin{tabular}{ l c c c c c  }
\hline\hline
\noalign{\smallskip}
Component & $v$ & $u$ &  $A_1$ & $A_2$ & $n$ \\
         & \multicolumn{2}{c}{(km\,s$^{-1}$)} &  &  & cm$^{-3}$ \\
\noalign{\smallskip}
\hline
\noalign{\smallskip}
  symmetric & 0     & 1800 &  0.59   & 0.27    & $1.25\times10^9$ \\
  red       & 1080  &  900 &  0.14   & 0.049   & $2.0\times10^8$ \\
  blue      & -450  &  480 &  0.32   & 0.20    & $2.8\times10^9$ \\
\noalign{\smallskip}
\hline
\end{tabular}
\end{table}
To address the disagreements between
   progenitor mass determinations, we should consider possible problems with our
   numerical modeling.
Among the missing factors that might affect the accuracy of the inferred SN
   parameters, the most apparent are: multi-group radiation transfer, full
   non-LTE treatment of gas excitation, time-dependent ionization, high
   spacial resolution of the shock front, multi-dimensional effects
   of the explosion and shock propagation, consideration of density
   perturbations related to vigorous convection in the RSG envelope.
The effects produced by these factors perhaps differ in magnitude and
   some of the factors could be insignificant.
However, detailed numerical studies based on advanced hydrodynamic
   models are needed to estimate the role of each factor.
At present, we can only state firmly that the multi-group treatment of
   radiation transfer cannot notably change the inferred SN parameters.
This conclusion is based on the comparison between
   the multi-group (Baklanov et al. \cite{BBP_05}) and one-group (Utrobin
   \cite{Utr_07}) approaches to the SN~1999em modeling.

A major drawback of our model may be its one-dimensional approximation.
The multi-dimensional effects related to the
   Rayleigh-Taylor mixing between the helium core and the hydrogen envelope
   during the shock propagation smear the composition and density jumps
   (M\"{u}ller et al. \cite{MFA_91}).
These effects are included artificially into our pre-SN model.
More careful treatment of these effects with multi-dimensional radiation
   hydrodynamics could modify the inferred SN parameters.

Another multi-dimensional effect, which could potentially be of importance,
   is the explosion asphericity.
A growing amount of observational data favor a picture in which the explosion of SNe~IIP
   is initiated by bipolar jets.
SN~1987A provided us with the first evidence of explosion asymmetry inferred from
   polarization (cf. Jeffery \cite{Jef_91}), line asymmetry (Haas et al.
   \cite{HCE_90}), and direct imaging (Wang et al. \cite{WWH_02}).
During the last decade polarization has been detected in another five SNe~IIP
   (Leonard \& Filippenko \cite{LF_01}; Leonard et al. \cite{LFA_01};
   Leonard et al. \cite{LFG_06}).
Two of these SNe~IIP, SN~1999em and SN~2004dj, exhibit pronounced asymmetry in
   their H$\alpha$ emission at the nebular stage, which is interpreted to be
   caused by asymmetric jets of $^{56}$Ni (Chugai \cite{Chu_07}).
The absence of a pronounced polarization at the early photospheric epoch
   indicates that the explosion asymmetry does not lead to the asphericity
   of the hydrogen envelope (Leonard et al. \cite{LFA_01}, \cite{LFG_06}).
The spherization would develop more successfully, if bipolar jets are
   thermal energy dominated (Couch et al. \cite{CWM_09}).
However, the spherization of jets in the hydrogen envelope does not preclude
   that the asymmetric explosion could result in the modification of the
   velocity-mass distribution compared to the spherical explosion.
The disagreement between the mass measurements may then be resolved, if
   the asymmetric explosion reproduces the observed light curve and
   expansion velocities for the essentially lower ejecta mass compared
   to the one-dimensional explosion model.
To verify this possibility, we would require multi-dimensional hydrodynamic modeling.

Using the spectra obtained by Sahu et al. (\cite{SASM_06}), we checked whether
   SN~2004et exhibited signatures of the explosion asymmetry.
We found that nebular H$\alpha$ and [O\,I] 6300, 6364 \AA\ lines
   indeed exhibit asymmetry.
To quantify asymmetry effects, each line of the oxygen doublet on day 301 was
   decomposed into three Gaussian components: symmetric, red, and blue.
The intensities of the corresponding components in blue and red lines of the
   doublet were free parameters.
Because the line optical depth affects the doublet ratio, as in SN~1987A
   (Chugai \cite{Chu_88}), the line ratio permits us to recover the Sobolev
   optical depth and, therefore, the oxygen number density.
The effect is weakly dependent on the electron temperature, which is assumed
   to be 5000~K.

\begin{figure}[t]
   \resizebox{\hsize}{!}{\includegraphics[clip, trim=0 0 0 240]{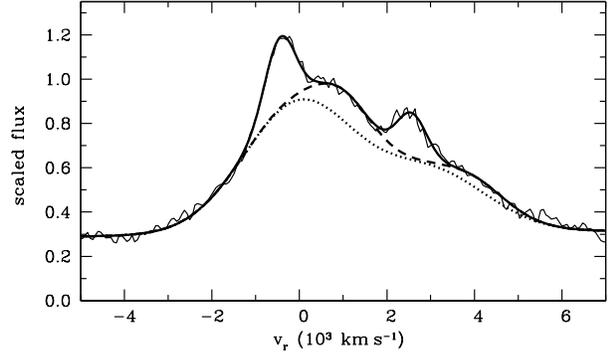}}
   \caption{%
   Oxygen doublet [O\,I] 6300, 6364 \AA\ observed on day 301 (\emph{thin solid
      line\/}).
   Zero radial velocity corresponds to the rest wavelength of 6300 \AA.
   \emph{Thick solid line} is the model doublet profile with all three Gaussian
      components, \emph{dashed line} is the profile without the blue component,
      and \emph{dotted line} is the symmetric component.
   }
   \label{fig:oxygen}
\end{figure}
The result of the doublet synthesis is shown in Fig.~\ref{fig:oxygen} with the
   model parameters listed in Table~\ref{tab:oxygen}.
The table columns give the radial velocity shift of the Gaussian component ($v$),
   its Doppler width ($u$), the amplitudes of its blue ($A_1$) and red ($A_2$)
   doublet components, and the number density of the line-emitting oxygen
   determined from the doublet ratio of each Gaussian component.
We emphasize that the decomposition is not unique unless we
   constrain ourselves by the shape and number of components.
The adopted decomposition procedure leads to a minimal contribution of
   the asymmetric components.
Although the red and blue components are weaker than the symmetric one,
   the asymmetry is rather pronounced.
Both asymmetric components have comparable integrated fluxes, but are not
   identical in terms of the Doppler widths and velocity shifts.
These results infer a bipolar structure and a deviation from the point
   symmetry of the line-emitting gas in the inner layers of the SN envelope.
We note our conclusion refers to the line-emitting gas, which is not identical to
   the overall oxygen distribution.
It may well be that the asymmetry of the line-emitting oxygen is related
   primarily to the asymmetry of $^{56}$Ni ejecta (Chugai \cite{Chu_07}).
Interestingly, a combination of the bipolar structure and the deviation from
   the point symmetry is a specific feature of SN~1987A, SN~1999em, and
   SN~2004dj (Chugai \cite{Chu_07}).
The case of SN 2004et thus provides further support to the conjecture that the
   explosion asymmetry is an ubiquitous phenomenon of SNe~IIP.

Remarkably, the oxygen number density for the symmetric and blue components
   (Table~\ref{tab:oxygen}) is comparable, to within a factor of unity, with
   the oxygen density of $(0.5-1.4)\times10^9$ cm$^{-3}$ for the optimal model
   in the velocity range $v<2500$ km\,s$^{-1}$.
In the latter case, we assume that the oxygen density is equal to the total
   matter density.
This coincidence suggests that whatever the role of asymmetry, the related
   effects do not strongly modify the density distribution of our optimal
   model in the inner layers with velocities $v<2000$ km\,s$^{-1}$.
Of course, this does not preclude that the velocity-density distribution
   in outer layers could be modified significantly as a result of
   the aspherical explosion.

\section{Conclusions}
\label{sec:concl}
Our goal was to recover the parameters of the hydrodynamic model of the luminous
   type IIP SN~2004et.
We obtained the optimal parameter set:
   the ejecta mass $M_{env}=22.9\pm1~M_{\sun}$, the explosion
   energy $E=(2.3\pm0.3)\times10^{51}$ erg, the pre-SN radius
   $R_0=1500\pm140~R_{\sun}$, and the $^{56}$Ni mass
   $M_{\mathrm{Ni}}=0.068\pm0.009~M_{\sun}$.
The inferred ejecta mass and explosion energy are maximal among all the known
   SNe~IIP explored by means of radiation hydrodynamics.
The parameters of SN~2004et strengthen correlations between the explosion
   energy and progenitor mass, and between the total $^{56}$Ni mass and
   progenitor mass discussed earlier (Utrobin \& Chugai \cite{UC_08}).

The progenitor mass of SN~2004et, estimated by combining the pre-SN mass and
   the mass lost via the stellar wind, turns out to be significantly, 
   by a factor of $2-3$, higher than the main-sequence mass recovered 
   from the pre-explosion images.
This and the disagreement between mass estimates found earlier for SN~2005cs
   raise serious concern about the reliability of the progenitor mass
   recovered from the hydrodynamic modeling.
We speculate that among the pitfalls of our hydrodynamic code, the most crucial
   could be the one-dimensional approximation.
The artificial mixing between the helium core and the hydrogen envelope, which
   we use to simulate real mixing, could be flawed, while explosion asphericity
   is completely ignored.
The evidence of the explosion asphericity of SN~2004et is inferred from the nebular
   lines, which thus supports the view that the explosion asymmetry is an
   ubiquitous phenomenon for SNe~IIP.

\begin{acknowledgements}
We are indebted to D. K. Sahu for sending us spectra of SN~2004et.
One of us (VU) is grateful to Wolfgang Hillebrandt for the possibility
   to work at the MPA.
We thank the anonymous referee for useful comments on our manuscript.
\end{acknowledgements}


\end{document}